\documentclass{iau}
\usepackage{graphicx}

\title[Morphologies of massive galaxies at $1 < z < 3$]
{The Morphologies of Massive Galaxies at ${\bf 1<z<3}$ in the CANDELS-UDS Field: Compact Bulges, and the Rise and Fall of Massive Disks}

\author[V.A. Bruce et al.]   
{V.A. Bruce$^{1}$\thanks{E-mail: vab@roe.ac.uk}, J.S. Dunlop$^{1}$, M. Cirasuolo$^{1,2}$, R.J. McLure$^{1}$, \\T.A. Targett$^{1}$, E.F. Bell$^{3}$, D.J. Croton$^{4}$, A. Dekel$^{5}$, S.M. Faber$^{6}$,\\ H.C. Ferguson$^{7}$, N.A. Grogin$^{7}$,  D.D. Kocevski$^{6}$, A.M. Koekemoer$^{7}$,  D.C. Koo$^{6}$, K. Lai$^{6}$, J.M. Lotz$^{7}$, E.J. McGrath$^{6}$, J.A. Newman$^{8}$,\\A. van der Wel$^{9}$}

\affiliation{$^1$SUPA\thanks{Scottish Universities Physics Alliance} Institute for Astronomy, University of Edinburgh, Royal Observatory, Edinburgh EH9 3HJ\\
$^2$UK Astronomy Technology Centre, Science and Technology Facilities Council, Royal Observatory, Edinburgh EH9 3HJ\\
$^3$Department of Astronomy, University of Michigan, 500 Church St., Ann Arbor, MI 48109, USA\\
$^4$Centre for Astrophysics and Supercomputing, Swinburne University of Technology, PO Box 218, Hawthorn, VIC 3122, Australia\\
$^5$Racah Institute of Physics, The Hebrew University, Jerusalem 91904, Israel\\
$^6$UCO/Lick Observatory, University of California, Santa Cruz, CA 95064, USA\\
$^7$Space Telescope Science Institute, 3700 San Martin Drive, Baltimore, MD 21218, USA\\
$^8$University of Pittsburgh, Pittsburgh, PA 15260, USA\\ 
$^9$Max-Planck Institut f\"{u}r Astronomie, K\"{o}nigstuhl 17, D-69117 Heidelberg, Germany\\  }

\pubyear{2013}
\volume{295}  
\pagerange{xx--xx}
\jname{The intriguing life of massive galaxies}
\editors{D. Thomas, A. Pasquali \& I. Ferreras, eds.}
\begin{document}

\maketitle

\begin{abstract}
We have used high-resolution, HST WFC3/IR, near-infrared imaging 
to conduct a detailed bulge-disk decomposition of the morphologies of $\simeq 200$ of the most massive ($M_* > 10^{11}\,{\rm M_{\odot}}$) galaxies at $1 < z < 3$ in the CANDELS-UDS field.
We find that, while such massive galaxies at low redshift are generally bulge-dominated, 
at redshifts $1<z<2$ they are predominantly mixed bulge+disk systems, and by $z > 2$ they are mostly disk-dominated. 
Interestingly, we find that while most of the quiescent galaxies are bulge-dominated, a significant fraction 
($25-40$\%) of the most quiescent galaxies, have disk-dominated morphologies.
Thus, our results suggest that the physical mechanisms which quench star-formation activity are not simply connected to those responsible for the morphological transformation of massive galaxies.

\keywords{galaxies: evolution - galaxies: structure - galaxies: spiral -  galaxies: elliptical and lenticular - cD,  galaxies: high-redshift}
\end{abstract}

\firstsection 
\section{Introduction}

The study of the high-redshift progenitors of today's massive galaxies can provide us with invaluable insights 
into the key mechanisms that shape the evolution of galaxies in the high-mass regime.

The latest generation of galaxy formation models
are now able to explain the number densities and ages of massive galaxies at high redshift. However, this is only part of the challenge, 
as recent studies have posed new questions about how the morphologies of massive galaxies evolve with redshift.

In addition to the basic question of how high-redshift galaxies evolve in size, there is also still much debate 
about how these massive galaxies evolve in terms of their fundamental morphological type. Extensive studies 
of the local Universe have revealed a bimodality in the colour-morphology plane, with spheroidal galaxies typically inhabiting 
the red sequence and disk galaxies making up the blue cloud (e.g. \cite[Baldry et al. 2004]{Baldry2004}). However, 
recent  studies at both low (e.g. \cite[Bamford et al. 2009]{Bamford2009}) and high redshift (e.g. \cite[van der Wel et al. 2011]{vanderWel2011}) have 
uncovered a significant population of passive disk-dominated galaxies, providing evidence that the physical processes 
which quench star-formation may be distinct from those responsible for driving morphological transformations. This result 
is particularly interesting in light of the latest morphological studies of high-redshift massive galaxies 
by \cite[Buitrago et al. (2013)] {Buitrago2013} and \cite[van der Wel et al. (2011)]{vanderWel2011} who find that, in contrast to the local population of massive galaxies
(which is dominated by bulge morphologies), by $z \simeq 2$ massive galaxies are predominantly disk-dominated systems. In this work we attempt to provide significantly improved clarity on these issues.

\section{Data and Morphological Fitting}

The CANDELS (\cite[Grogin et al. 2011]{Grogin2011}, \cite[Koekemoer et al. 2011]{Koekemoer2011}) near-infrared F160W data provides the necessary combination of depth, angular resolution, and area to enable the most detailed study to date of the rest-frame optical morphologies of massive ($M_* > 10^{11}\,{\rm M_{\odot}}$) galaxies at $1 < z < 3$ in the UKIDSS Ultra Deep Survey (\cite[Lawrence et al. 2007]{Lawrence2007}). 
For this study we have constructed a sample based on photometric redshifts and stellar mass estimates which were determined using 
the stellar population synthesis models of \cite[Bruzual \& Charlot (2003)]{Bruzual2003} assuming a Chabrier initial mass function (see \cite[Bruce et al. 2012]{Bruce2012} for full details).
This provides us with a total mass-complete sample of $\sim200$ galaxies.

We have employed the GALFIT (\cite[Peng et al. 2002]{Peng2002}) morphology fitting code 
to determine the morphological properties for all the objects in our sample. To conduct the double component fitting we define three components: a S\'{e}rsic index fixed at $n=4$ bulge, an $n=1$ fixed disk and a centrally concentrated PSF component to account for any AGN or nuclear starbursts within our galaxies. These three components are combined to generate six alternative multiple component model fits, of varying complexity, for every object in the sample. These models are formally nested, and thus $\chi^{2}$ statistics can be used 
to determine the ``best'' model given the appropriate number of model parameters.

\section{Evolution of Morphological Fractions}

 Armed with this unparalleled morphological
 information on massive galaxies at high redshift we can consider how the relative number density of galaxies of different morphological type
changes during the key epoch in cosmic history probed here. 

In Fig.~1 we illustrate this by binning our sample into four 
redshift bins of width $\Delta z = 0.5$, and consider three alternative cuts in morphological classification
as measured by $B/T$ from our bulge-disk decompositions.  In the left-hand panel
of Fig.~1 we have simply split the sample into two categories: bulge-dominated ($B/T > 0.5$) and disk-dominated ($B/T < 0.5$). In the central
panel we have separated the sample into three categories, with any object for which $0.3 < B/T < 0.7$ classed as ``Intermediate''.
Finally, in the right-hand panel we have expanded this Intermediate category to encompass all objects for which $0.1 < B/T < 0.9$.

\begin{figure}[h]
\begin{center}
 \includegraphics[width=5.5in]{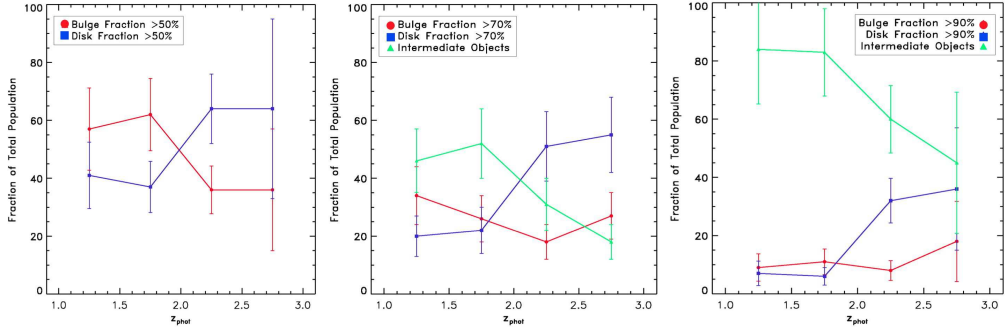} 
 \caption{The redshift evolution of the morphological fractions in our galaxy sample, after binning into 
redshift bins of width $\Delta z=0.5$ and using three alternative cuts in morphological classification (both to try to provide 
a complete picture, and to facilitate comparison with different categorisations in the literature).}
   \label{fig1}
\end{center}
\end{figure}

From these panels it can be seen that $z \simeq 2$ marks a key transition phase, above which massive galaxies are predominantly disk-dominated
 systems and below which they become increasingly mixed bulge+disk systems. We also note that at the lowest redshifts probed by this study ($z\simeq1$) 
 it is seen that, while bulge-dominated objects are on the rise,  
pure-bulge galaxies (i.e. objects comparable to present-day giant ellipticals) have yet to emerge in significant numbers, with $>90$\%
of these high-mass galaxies still retaining a significant disk component. This is compared with $64\%$ of the local $M_* > 10^{11}\,{\rm M_{\odot}}$ 
galaxy population, which would be classified as pure-bulges from our definition ($B/T>0.9$, corresponding to $n>3.5$ ) from the sample of \cite[Buitrago et al. (2013)]{Buitrago2013}.
Thus, our results further challenge theoretical models of galaxy formation to account for the relatively
rapid demise of massive star-forming disks, but the relatively gradual emergence of genuinely bulge-dominated morphologies.

\section{Star-forming and Passive Disks.}

In addition to our morphological decompositions we also make use of the SED fitting already employed in the sample selection to explore 
the relationship between star-formation activity and morphological type. 

Fig. 2 shows specific star-formation rate ($sSFR$) versus morphological type for the massive galaxies in our sample, where morphology 
is quantified  by single S\'{e}rsic index in the left-hand panel, and by bulge-to-total $H_{160}$-band flux ratio ($B/T$) in the right-hand panel.
The values of $sSFR$ plotted are derived from the original optical-infrared SED fits employed in the sample selection, and include
correction for dust extinction as assessed from the best fitting value of $A_V$ derived
during the SED fitting. As a check of the potential failure of this approach to correctly identify reddened dusty star-forming 
galaxies,  we have also searched for 24\,$\mu$m counterparts in the {\it Spitzer} SpUDS MIPS imaging of the UDS, 
and have highlighted in blue stars those objects which yielded a MIPS counterpart within a search radius of $<2$\,arcsec. 

To first order, our results show that the well-documented bimodality in the
colour-morphology plane seen at low redshift, where spheroidal
galaxies inhabit the red sequence, while disk galaxies occupy the blue cloud
is at least partly already in place by $z \simeq 2$.

Nonetheless, the sample
also undoubtedly contains star-forming bulge-dominated galaxies and, perhaps more interestingly, a significant 
population of apparently quiescent disk-dominated objects. To highlight and quantify  
this population we have indicated by a box on both the panels the region occupied by objects with disk-dominated morphologies and 
$sSFR < 10^{-10}\,{\rm yr^{-1}}$. In the left-hand panel, disk-dominated is defined as $n < 2.5$, and $40 \pm 7$\% of the quiescent galaxies 
lie within this box (if we exclude the 24\,$\mu$m detections), 
while in the right-hand panel, disk-dominated is defined by $B/T < 0.5$, 
in which case $25 \pm 6$\% of the quiescent objects lie within this region.

 \begin{figure}[h]
\begin{center}
 \includegraphics[width=4.5in]{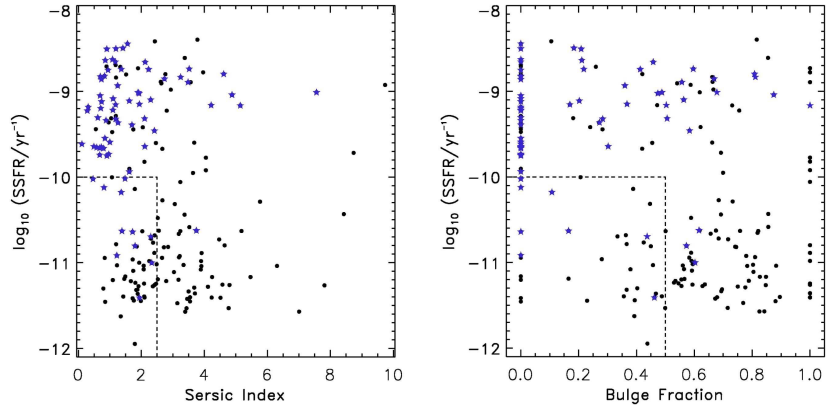} 
 \caption{Plots of specific star-formation rate ($sSFR$) versus morphological type as 
judged by single S\'{e}rsic index (left-hand panel) and bulge-to-total $H_{160}$-band flux ratio ($B/T$) (right-hand panel).}
   \label{fig1}
\end{center}
\end{figure}

The presence of a significant population of passive disks among the massive galaxy population
at these redshifts indicates that star-formation activity can cease without a disk galaxy
being turned directly into a disk-free spheroid, as generally previously expected if the process that
quenches star formation is a major merger. 

One possible mechanism for this arises from the latest generation of hydrodynamical simulations (e.g. \cite[Kere\v{s} et al. 2005]{Keres2005}, \cite[Dekel et al. 2009a]{Dekel2009a}) 
and analytic theories (e.g. \cite[Birnboim \& Dekel 2003]{Birnboim2003}), which suggest a formation scenario whereby at high redshift star-formation is fed through inflows of cold gas. 
Another scenario which can account for star-formation quenching, whilst still being consistent with 
the existence of passive disks, is the model of violent disk instabilities 
(e.g. \cite[Dekel et al. 2009b]{Dekel2009b}), coupled with ``morphology quenching" (\cite[Martig et al. 2009]{Martig2009}).

\end{document}